\documentclass{article}
\pdfoutput=1
\usepackage{enumitem}
\usepackage{vmargin}
\setpapersize{USletter}
\setmarginsrb{1in}{1in}{1in}{1in}{0pt}{0pt}{0pt}{12pt}

\usepackage{latexsym}
\usepackage{amssymb}
\usepackage{graphics}
\usepackage{url}

\newcommand{\intring}{\mathbb{Z}}

\newcommand{\realring}{\mathbb{R}}

\newtheorem{definition}{Definition}
\newtheorem{theorem}{Theorem}
\newtheorem{lemma}{Lemma}
\newtheorem{corollary}{Corollary}

\newcommand{\true}{\textsc{True}}
\newcommand{\false}{\textsc{False}}
\newcommand{\fail}{\textsc{Fail}}

\newcommand{\myand}{\wedge}

\newcommand{\ldcf}[2]{\mbox{ldcf}_{#1}(#2)}
\newcommand{\factors}[1]{\mbox{factors$\left(#1\right)$}}
\newcommand{\sqr}[2]{{\vcenter{\hrule height.#2pt
        \hbox{\vrule width.#2pt height#1pt \kern#1pt
           \vrule width.#2pt}
        \hrule height.#2pt}}}
 
\newcommand{\proof}{\noindent{\sc Proof.}\enspace}

\newcommand{\qed}{{\nobreak\hfil\penalty50
  \hskip1em\hbox{}\nobreak\hfil$\sqr{8}{8}$
  \parfillskip=0pt \finalhyphendemerits=0 \par}
  \bigskip}

\begin{document}

\title{Model-based construction of Open Non-uniform Cylindrical
  Algebraic Decompositions}
\author{Christopher W. Brown\\
United States Naval Academy\\
Annapolis, Maryland 21402\\
wcbrown@usna.edu}
\maketitle

\begin{abstract}
In this paper we introduce the notion of an Open
\emph{Non-uniform Cylindrical Algebraic Decomposition} (NuCAD),
and present an efficient \emph{model-based} algorithm for 
constructing an Open NuCAD from an input formula.
A NuCAD is a generalization of Cylindrical Algebraic Decomposition
(CAD) as defined by Collins in his seminal work from the early 
1970s, and as extended in concepts like Hong's partial CAD.
A NuCAD, like a CAD, is a decomposition of $\realring^n$ into 
cylindrical cells.  But unlike a CAD, the cells in a NuCAD need not
be arranged cylindrically.  It is in this sense that NuCADs are
not uniformly cylindrical.
However, NuCADs --- like CADs --- carry a tree-like structure that
relates different cells.  It is a very different tree but, as with
the CAD tree structure, it allows some operations to be performed
efficiently, for example locating the
containing cell for an arbitrary input point.
\end{abstract}

\section{Introduction}
This paper introduces a new \emph{model-based} approach to 
constructing Cylindrical Algebraic Decompositions (CADs).  The
model-based approach, building on \cite{JovanovicdeMoura:12}
and \cite{Brown:2013}, has some very nice properties 
(described later in the paper) that make it appealing.
However, prior work has not applied it to constructing CADs.
Jovanovic and de Moura's work \cite{JovanovicdeMoura:12}, which
introduced the approach, uses it to determine the satisfiability of
Tarski formulas.  In some sense, their approach can be seen as
building a CAD-like decomposition.  However, what is constructed is
an unstructured list of cells, which makes it unsuitable for some
of what CADs are used for.  Moreover, the method is not obviously
parallelizable, and it doesn't take as strong advantage of the 
``model-based approach'' as is possible. \cite{Brown:2013} shows
how to make stronger use of the ``model'' during the construction
of a single open cell.  This paper continues in one of the directions
outlined in that paper, using the strong model-based approach to 
construct not just a single cylindrical cell, but a whole
decomposition of real space into cylindrical cells.  

A particularly
exciting aspect of this new model-based approach is that while each 
cell in the
decomposition is cylindrical, those cells need not by cylindrically 
arranged with respect to one another.  This frees us to construct 
more general decompositions than CADs, thereby representing
semi-algebraic sets with fewer cells.  To make use of this freedom, we
introduce a new generalization of CAD, the 
\emph{Open Non-uniform Cylindrical Algebraic Decomposition} (Open NuCAD),
and an algorithm TI-Open-NuCAD that efficiently constructs an Open
NuCAD from an input formula.  As demonstrated by an example
computation that is worked out in detail in this paper, the
flexibility of NuCADs allow sets to be represented using fewer cells 
than with a CAD.

\section{Non-uniform Cylindrical Algebraic Decomposition}
In this section we define Non-uniform Cylindrical Algebraic
Decomposition.  We assume the reader is already familiar with
the usual CAD notions --- like delineability, level of a polynomial,
etc.  Note that $\lambda$ denotes the empty string in what follows,
$||$ indicates concatenation,
and $\pi_k(\cdot)$ denotes projection down onto $\realring^k$.
This paper deals with open cylindrical cells which, except in the
trivial case of a single cell, cannot truly decompose $\realring^n$.
Instead, we say that a set of open regions defines a 
\emph{weak decomposition} of $\realring^n$ if the regions are 
pairwise disjoint, and the union of their closures contains $\realring^n$.
We here provide a definition of an open cylindrical cell.  This is
entirely in keeping with the usual definition of a cell in the CAD
literature.

\begin{definition}
\label{definition:NuCAD}
An \emph{Open Cylindrical Cell} is a subset of $\realring^n$
is a set of the form 
$$
\{ (\alpha_1,\ldots,\alpha_n) \in B\times\realring | 
f(\alpha_1,\ldots,\alpha_{n-1}) < \alpha_n < g(\alpha_1,\ldots,\alpha_{n-1})\} 
$$
or
$$
\{ (\alpha_1,\ldots,\alpha_n) \in B\times\realring | 
f(\alpha_1,\ldots,\alpha_{n-1}) < \alpha_n\} 
$$
or
$$
\{ (\alpha_1,\ldots,\alpha_n) \in B\times\realring | 
\alpha_n < g(\alpha_1,\ldots,\alpha_{n-1})\} 
$$
where $B$ is an open cylindrical cell in $\realring^{n-1}$ and
the graphs of $f$ and $g$ over $B$ are disjoint sections of
polynomials,
and $()$ is considered an open cylindrical cell in $\realring^0$.
\end{definition}

Next we define Open Non-uniform Cylindrical Algebraic Decomposition
(Open NuCAD), which relaxes the requirements of the usual CAD.  In
particular, it is possible to have two cells whose projections onto
a lower dimension are neither equal nor disjoint.  In other words,
while each individual cell is cylindrical, distinct cells are not
necessarily organized into cylinders.

\begin{definition}
  \label{definition:OpenNuCAD}
  An \emph{Open Non-uniform Cylindrical Algebraic Decomposition} 
  (Open NuCAD) of $\realring^n$ 
  is a collection $C$ of open cylindrical cells, each of
  which is labelled with a unique string of the form 
  $([0-9]+(L,U,X))*$. 
  The relation 
  $$
  E = \{(C_1,C_2) | \mbox{$C_1$ and $C_2$ are cells with labels
    $lab_1$ and $lab_2$ 
    satisfying $lab_2 = lab_1([0-9]+(L,U,X))$}\}$$
    defines a graph on the cells.
  \begin{enumerate}
    \item the graph $(C,E)$ is a tree, rooted at cell $\realring^n$,
      with label $\lambda$ (the empty string),
    \item the children of cell $C_0$ with label  $lab_0$ have labels
      taken from the set
      $$\{lab_01L,\ldots,lab_0nL,lab_01U,\ldots,lab_0nU,lab_0nX\}$$
      and if $C_0$ has children, then one of them is labelled 
      $lab_0nX$,
    \item if cell $C_2$ is the child of $C_1$ with label
      $lab_1nX$, then $C_2 \subseteq  C_1$ and for each $i \in \{1,\ldots,n\}$,
      in the cylinder over over $\pi_{i-1}(C_2)$
      the section that defines the lower (resp. upper)
      boundary of $C_2$ in $x_i$ is either identical to or disjoint 
      from the section that defines the lower (resp. upper)
      boundary of $C_1$ in $x_i$ 
    \item 
      \label{point:LandU}
      if cell $C_X$ is the child of $C_0$ with label
      $lab_0nX$, 
      then 
      \begin{equation}
        \label{equation:ExprLandU}
      (\pi_{i-1}(C_X) \times \realring) \cap \pi_i(C_0) - \overline{\pi_i(C_X)}
      \end{equation}
      consists of zero one or two open cells: the region with
      $i$-coordinates below
      $\overline{\pi_i(C_X)}$ if it is non-empty, which is denoted
      $B_L$,
      and the region with $i$-coordinates above
      $\overline{\pi_i(C_X)}$ if it is non-empty, which is denoted
      $B_U$. 
      There is a cell with label $lab_0iL$ if and only if
      $B_L$ is non-empty and, if it exists, that cell is
      $(B_L \times \realring^{n-i}) \cap C_0$.
      There is a cell with label $lab_0iU$ if and only if
      $B_U$ is non-empty and, if it exists, that cell is
      $(B_U \times \realring^{n-i}) \cap C_0$.
  \end{enumerate}
\end{definition}

Next we prove that NuCADs really do define decompositions of
$\realring^n$ or, more properly, Open NuCADs define weak
decompositions of $\realring^n$.

\begin{theorem}
If cell $C_0$ is a non-leaf node in the graph $(C,E)$, 
its children form a weak decomposition of $C_0$.
\end{theorem}
\proof
What needs to be proved is that there is no open subset of $C_0$
having empty intersection with all of the children of $C_0$.
Let $S$ be an open, connected subset of $C_0$.  Let $i$ be the maximum
element of $\{1,\ldots,n+1\}$ such that $\pi_{i-1}(S) \subseteq \pi_{i-1}(C_X)$.  
If $i = n+1$, then $S$ is contained in $C_X$, the child that, by
definition, must exist.  So the theorem holds in this case.

If $i \leq n$, then we have $\pi_{i-1}(S) \subseteq \pi_{i-1}(C_X)$, but
$\pi_{i}(S) \nsubseteq \pi_{i}(C_X)$.  Consider the key expression (\ref{equation:ExprLandU})
from Point~\ref{point:LandU} of Definition~\ref{definition:OpenNuCAD} with
regards to $i$:
\[
\underbrace{
(\overbrace{\pi_{i-1}(C_X)}^{\pi_{i-1}(S)\subseteq} \times \realring)
\cap \pi_i(C_0)}_{\pi_{i}(S)\subseteq} -
\underbrace{\overline{\pi_i(C_X)}}_{\pi_{i}(S)\nsubseteq}
\]
This shows that one or both of the regions $B_L$ and $B_U$ from Point~\ref{point:LandU}
have non-empty intersection with $\pi_i(S)$, and thus is/are
non-empty.  Suppose $B_L \cap \pi_i(S) \neq \emptyset$ (the case for
$B_U$ is entirely analogous, and so will not be given explicitly).
Since $B_L$ is non-empty, by definition
$C_0$ has a child with label $lab_0iL$ that is $(B_L \times \realring^{n-i}) \cap C_0$.
Since $S\subseteq C_0$ and $\pi_i(S) \subseteq B_L$, we have
$$\left((B_L \times \realring^{n-i}) \cap C_0\right) \cap S \neq \emptyset,$$
which proves the theorem.
\qed

\begin{corollary}
The leaf cells of an Open NuCAD comprise a weak decomposition of $\realring^n$.
\end{corollary}

\section{Algorithms}

We will follow the OpenCell data structure definition 
provided in \cite{Brown:2013}, with the following 
additions:
\begin{enumerate}
\item each cell carries a sample point $\alpha$ with it
\item each cell has an associated set $P$ of irreducible
  polynomials that are known to be sign-invariant 
  (which implies order-invariant, since these are open cells)
  within the cell.
\item each cell has an associated label $lab$ of the form
  $([0-9]+(L,U,X))*$.
\end{enumerate}

We assume the existence of a procedure OC-Merge-Set
that is analogous to the procedure O-P-Merge defined in
\cite{Brown:2013}, except that instead of merging a 
single polynomial $P$ with a given OneCell $C$, it merges 
a set $Q$ of polynomials with a given OneCell $C$.  This
could be realized by simply applying O-P-Merge iteratively, 
or via a divide-and-conquer approach as alluded to in the 
final section of \cite{Brown:2013}.  We will assume that
this procedure manipulates OneCell data structures with the
augmentations described above.  The label $lab$ and point 
$\alpha$ for the refined cell returned by OC-Merge-Set is
simply inherited from the input OneCell $C$, and the 
associated set of polynomials is the super-set of $P \cup Q$
(where $P$ is the set associated with $C$) defined by the 
projection factors computed during the refinement process
--- all of which are known to be sign-invariant in the refined
OneCell.

\textbf{Algorithm: Split}\\
\textbf{Input:}
OpenCell $D$ (with point $\alpha \in \realring^n$,
projection factor set $P$, and label $lab$), and Formula $F$\\
\textbf{Output:}
queue of OpenCells that is either empty (in which case $F$ is
truth-invariant in $D$), or whose elements comprise a valid set
of children for $D$ according to Definition~\ref{definition:NuCAD}
(in which case $F$ is truth-invariant in the cell with label $labnX$).
\begin{enumerate}
\item choose $Q \subset \intring[x_1,\ldots,x_n]$ such that
  $Q \cap P = \emptyset$ and the sign-invariance of the elements of $P \cup Q$
  within a connected region containing $\alpha$ implies the
  truth-invariance of $F$; if $Q = \emptyset$ return an empty queue

\item $D' = \mbox{OC-Merge-Set}(D,\alpha,Q)$

\item if $D' = (\fail,f)$ then \emph{/$\ast$ perturb $\alpha$ $\ast$/}
  \begin{enumerate}
  \item $L = \{f\}$, $i = \mbox{ level of $f$}$
  \item while at least one element of $L$ is nullified at
    $(\alpha_1,\ldots,\alpha_{i-1})$ do
    \begin{enumerate}
      \item $L = \bigcup_{g \in L} \factors{\ldcf{x_i}{g}}$
      \item $i = i - 1$
    \end{enumerate}
  \item $\zeta = \mbox{max}\{ \beta\in \realring\ |\ \beta < \alpha_i \mbox{ and }
    g(\alpha_1,\ldots,\alpha_{i-1},\beta) = 0 \mbox{ for some $g\in L$}\}$
  \item choose $\gamma_i \in \left(\max(\zeta,D[i].L),\alpha_i\right)$
  \item for $j$ from $i+1$ to $n$ do
    \begin{enumerate}
      \item choose $\gamma_j$ so that
        \[
        \begin{array}{c}
          root(D[j].l(\alpha_1,\ldots,\alpha_{i-1},\gamma_i,\ldots,\gamma_{j-1},x_j),
          D[j].L.j,x_j) < \gamma_j \mbox{ and}\\
          \gamma_j < 
          root(D[j].u(\alpha_1,\ldots,\alpha_{i-1},\gamma_i,\ldots,\gamma_{j-1},x_j),
          D[j].U.j,x_j)
        \end{array}
        \]
      \end{enumerate}
  \item set $\alpha =
    (\alpha_1,\ldots,\alpha_{i-1},\gamma_i,\ldots,\gamma_n)$,
    adjusting data-structure $D$ accordingly
  \item goto Step~2
  \end{enumerate}
  
\item enqueue $D',\alpha,P',lab'$ on output queue, where $P'$ is
  produced by the merge process, and $lab' = lab||nX$

\item for $i$ from 1 to $n$ do \emph{/$\ast$ split $D$ based on $D'$ $\ast$/}
  
      \begin{enumerate}
      \item 
        if $D'[i].l \neq D[i].l$ then \emph{/$\ast$ lower bound at
          level $i$ changes $\ast$/}
        \begin{enumerate}
        \item $D'_{iL} = D'[1],\ldots,D'[i-1],
          (D[i].l,D[i].L,D'[i].l,D'[i].L)
          ,D[i+1],\ldots,D[n]$
        \item for $j$ from $i$ to $n$, choose $\gamma_j$ so that
          \[
          \begin{array}{c}
            root(D'_{iL}[j].l(\alpha_1,\ldots,\alpha_{i-1},\gamma_i,\ldots,\gamma_{j-1},x_j),
            D'_{iL}[j].L.j,x_j) < \gamma_j \mbox{ and}\\
            \gamma_j < 
            root(D'_{iL}[j].u(\alpha_1,\ldots,\alpha_{i-1},\gamma_i,\ldots,\gamma_{j-1},x_j),
            D'_{iL}[j].U.j,x_j)
          \end{array}
          \]
        \item $\alpha'_{iL} = (\alpha_1,\ldots,\alpha_{i-1},\gamma_i,\ldots,\gamma_n)$
          \item $P'_{iL} = P \cup (P' \cap
            \realring[x_1,\ldots,x_{i-1}])$, where $P$ and $P'$ are
            the sign-invariant polynomial sets for $D$ and $D'$, respectively.
            {\small Note: we might sometimes deduce that there are
              other polynomials that are sign-invariant in $D'_{iL}$.
            This could be quite worthwhile!}
        \item $lab_{iL} = lab||iL$, where $lab$ is the label for $D$
        \item enqueue new cell $D'_{iL},\alpha'_{iL},P'_{iL},lab_{iL}$
          in output queue
        \end{enumerate}
      
      \item 
        if $D'[i].u \neq D[i].u$ then \emph{/$\ast$ upper bound at
          level $i$ changes $\ast$/}
        \begin{enumerate}
        \item $D'_{iU} = D'[1],\ldots,D'[i-1],
          (D[i]'.u,D'[i].U,D[i].u,D[i].U)
          ,D[i+1],\ldots,D[n]$
        \item for $j$ from $i$ to $n$, choose $\gamma_j$ so that
          \[
          \begin{array}{c}
            root(D'_{iU}[j].l(\alpha_1,\ldots,\alpha_{i-1},\gamma_i,\ldots,\gamma_{j-1},x_j),
            D'_{iU}[j].L.j,x_j) < \gamma_j \mbox{ and}\\
            \gamma_j < 
            root(D'_{iU}[j].u(\alpha_1,\ldots,\alpha_{i-1},\gamma_i,\ldots,\gamma_{j-1},x_j),
            D'_{iU}[j].U.j,x_j)
          \end{array}
          \]
          \item $\alpha'_{iU} = (\alpha_1,\ldots,\alpha_{i-1},\gamma_i,\ldots,\gamma_n)$
          \item $P'_{iU} = P \cup (P' \cap
            \realring[x_1,\ldots,x_{i-1}])$, where $P$ and $P'$ are
            the sign-invariant polynomial sets for $D$ and $D'$, respectively.
            {\small Note: we might sometimes deduce that there are
              other polynomials that are sign-invariant in $D'_{iL}$.
            This could be quite worthwhile!}
        \item $lab_{iU} = lab||iU$, where $lab$ is the label for $D$
        \item enqueue new cell $D'_{iU},\alpha'_{iU},P'_{iU},lab_{iU}$
          in output queue
        \end{enumerate}
      \end{enumerate}
    \item return output queue
\end{enumerate}

\textbf{Algorithm: TI-Open-NuCAD}\\
\textbf{Input:} Formula $F$ in variables $x_1,\ldots, x_n$\\
\textbf{Output:}
Open-NuCAD $C$ in the leaf cells of which $F$ is truth invariant
\begin{enumerate}
  \item $C = \{ \}$
  \item let $Q$ be an empty queue
  \item enqueue in $Q$ and add to $C$ the OneCell representing $\realring^n$, with point
    $\alpha$ chosen arbitrarily, $P = \{ \}$, and label $lab =
    \lambda$.
  \item while $Q$ is not empty
    \begin{enumerate}
      \item dequeue $D$ from $Q$ \emph{/$\ast$ need not actually
          follow FIFO $\ast$/}
      \item if the label of $D$ ends in $X$, continue to next iteration
      \item $Q' = \mbox{Split}(D,F)$
      \item for each $D'$ in $Q'$ do
        \begin{enumerate}
        \item add $D'$ to $Q$
        \item add $D'$ to $C$
        \end{enumerate}
    \end{enumerate}
  \item return $C$
\end{enumerate}

Note that no one method for choosing $Q$ in Step 1 of the algorithm
Split is specified.  There are different ways to do this, and which
one is employed may well affect practical performance quite a bit and
will warrant future investigation.  One point we will make, however,
is that $\alpha$ plays a role in making this choice.  For example,
suppose $F = f_1 > 0 \myand f_2 > 0 \myand \cdots \myand f_r > 0$, and
suppose $F$ is $\false$ at $\alpha$.
To choose $Q$ we need only find one $f_i \notin P$ that is negative at
$\alpha$. If there are multiple such $f_i$'s, we could choose among
them in several different ways.  We could take the lowest level $f_i$.
We could prefer low-degree $f_i$'s.  If all potential $f_i$'s are of
level $n$, we could substitute $(\alpha_1,\ldots,\alpha_{n-1})$ into
all of them, examine the CAD of $\realring^1$ that results, and choose
the $f_i$ based on that information.

\section{An Example Open NuCAD Construction}
Consider the input formula 
$F = [ 16 y - 16 x^2 - 8 x - 1 > 0 \myand x^2 + y^2 - 1 > 0 ]$.
We will follow the execution Algorithm TI-Open-NuCAD on this input.
In the interest of space, we will name the polynomials that will
appear in the computation up front:
$$
f_1 = 16 y - 16 x^2 - 8 x - 1 , f_2 = x^2 + y^2 - 1,
f_3 = 256 x^4 + 256 x^3 + 352 x^2 + 16 x - 255, f_4 = x+1, f_5 = x-1
$$
\pagebreak
\begin{enumerate}
\item Cell $C_0 = ([\ ],lab=\lambda, \alpha=(0,0), P=\{\})$ consisting of $\realring^2$ enqueued on $Q$
\raisebox{-.4\height}{\scalebox{0.1}{\includegraphics{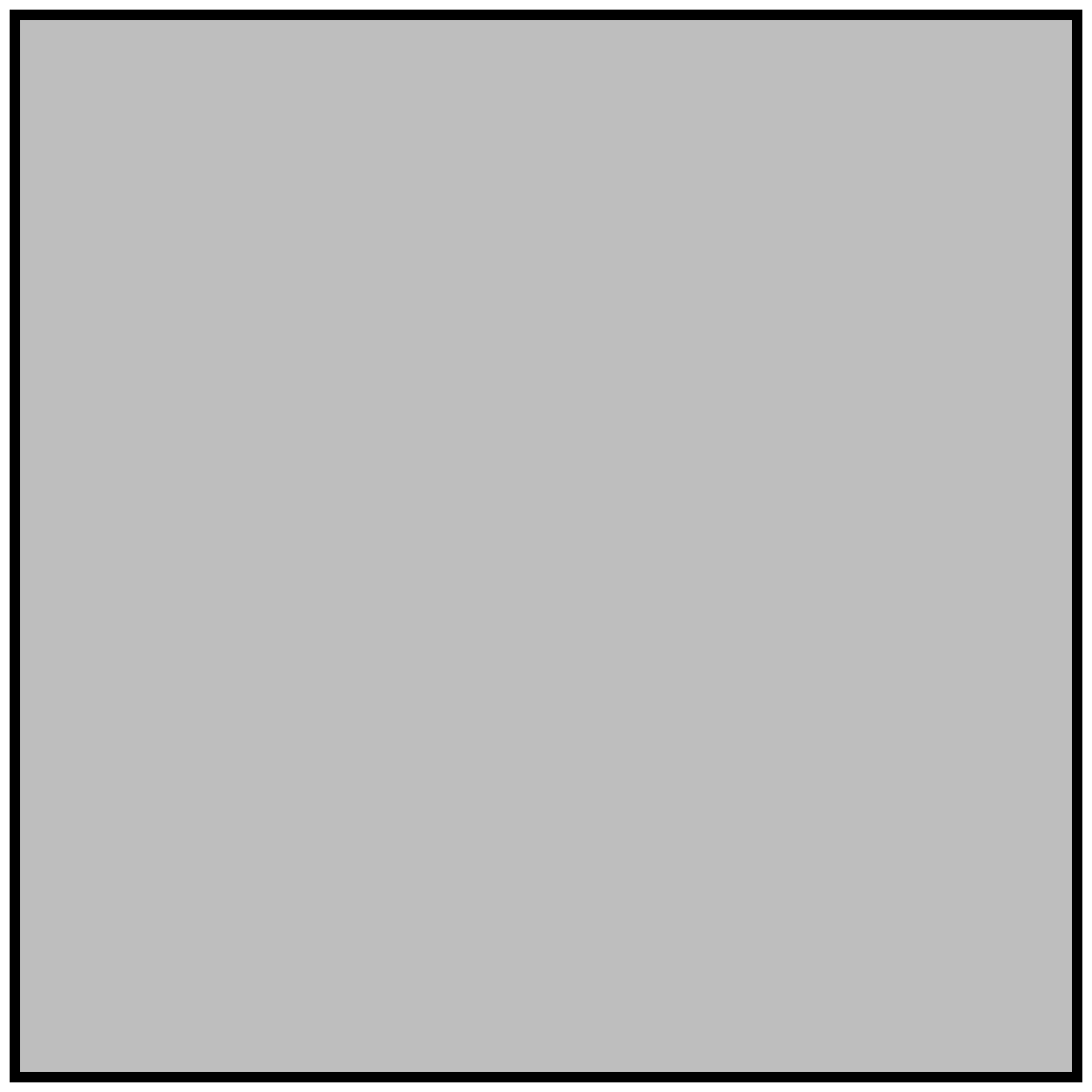}}}

\item Split($C_0$): $F(\alpha)=\false$, choose $Q = \{f_1\}$, enqueue
  the following
  cells
  \begin{itemize}[leftmargin=0in]
  \item $C_1 = ([y<root(f_1,1,y)],lab=2X,\alpha=(0,0),P={f_1})$ \raisebox{-.4\height}{\scalebox{0.1}{\includegraphics{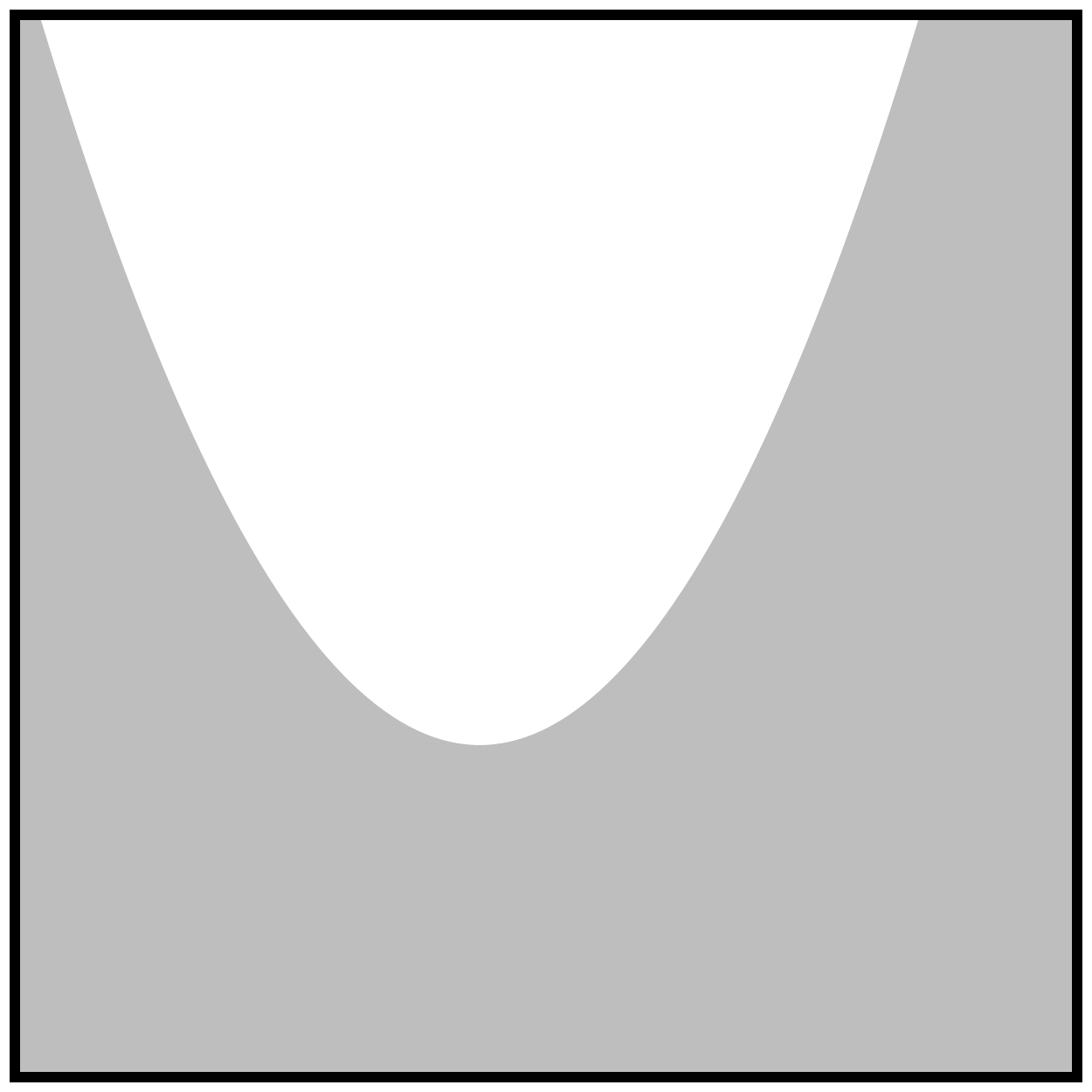}}}
  \item $C_2 = ([y>root(f_1,1,y)],lab=2U,\alpha=(0,1/2),P={f_1})$ \raisebox{-.4\height}{\scalebox{0.1}{\includegraphics{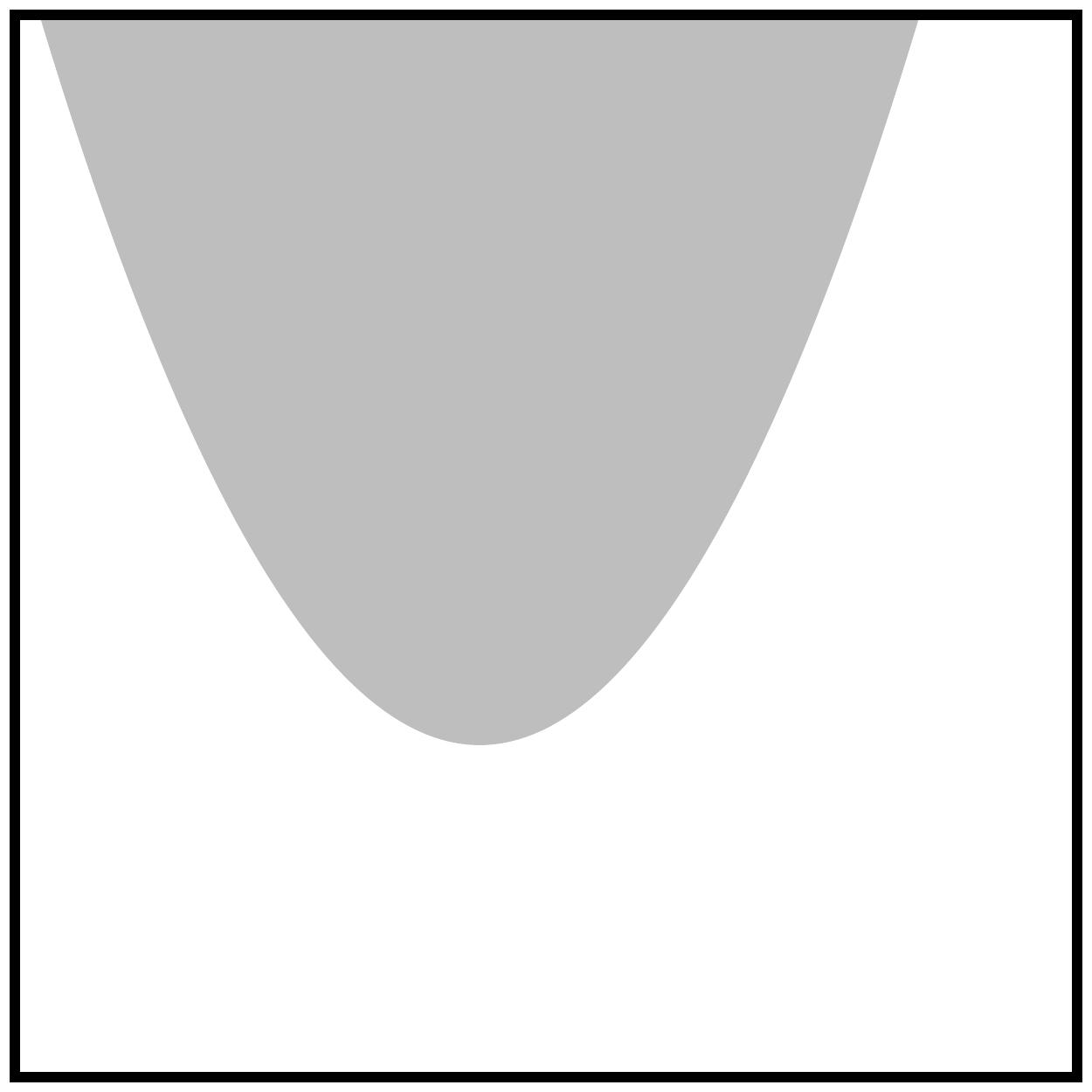}}}
  \end{itemize}
  \item $C_1$'s label ends in $X$, so it is not processed further
  \item Split($C_2$): $F(\alpha)=\false$, choose $Q = \{f_2\}$, enqueue
  the following
  cells
   \begin{itemize}[leftmargin=0in]
   \item $C_3 = \left(\left[
       \begin{array}{c}
         y>root(f_1,1,y)\myand y < root(f_2,2,y)\myand \\
         x > root(f_3,1,x) \myand x < root(f_3,2,x)
         \end{array}
       \right],
lab=2U2X,\alpha=(0,\frac{1}{2}),P=\{f_1,f_2,f_3,f_4,f_5\}\right)$ \raisebox{-.4\height}{\scalebox{0.1}{\includegraphics{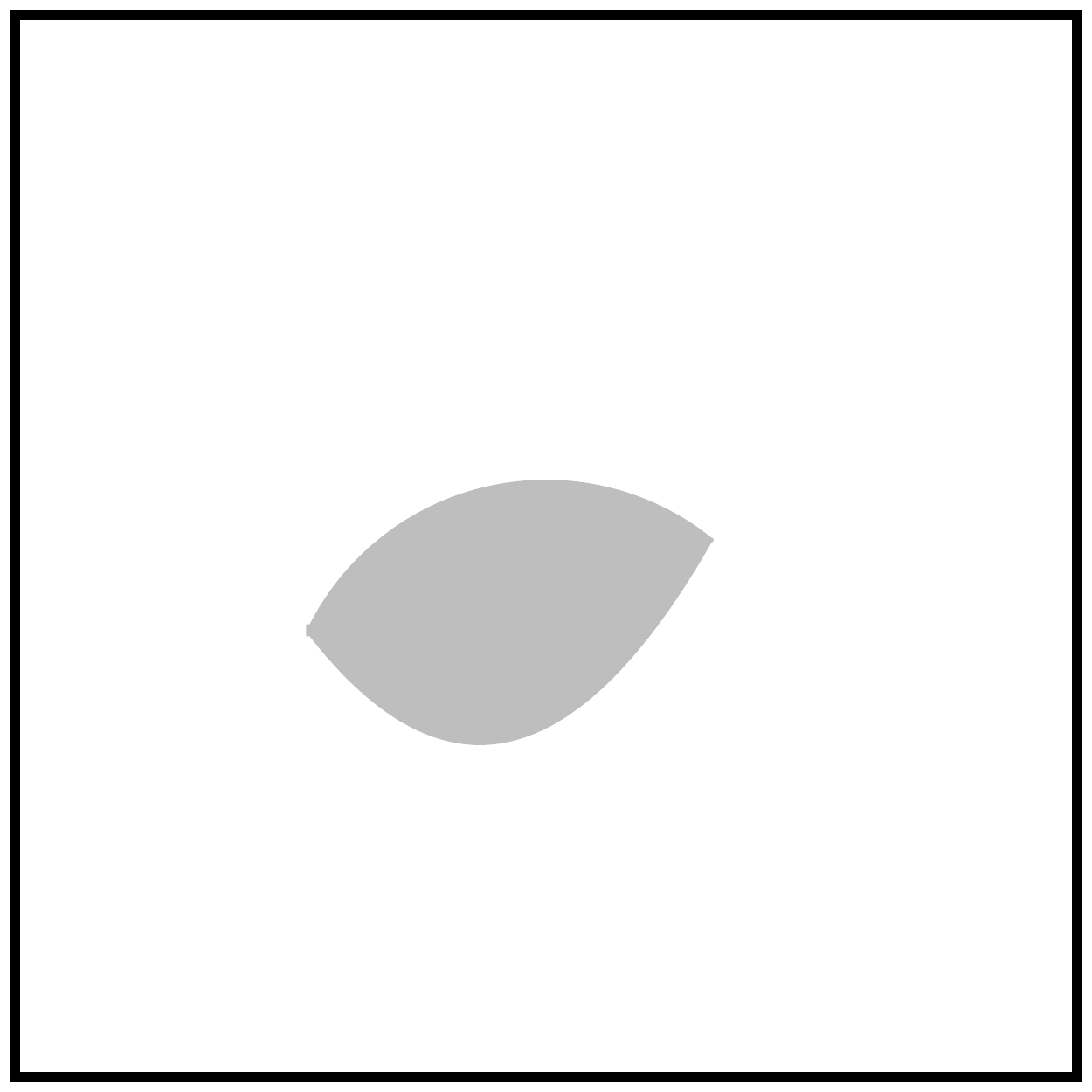}}}
   \item $C_4 = \left(\left[
       \begin{array}{c}
         y>root(f_1,1,y)\myand x < root(f_3,1,x)
         \end{array}
       \right],
lab=2U1L,\alpha=(-\frac{3}{2},2),P=\{f_1,f_3\}\right)$ \raisebox{-.4\height}{\scalebox{0.1}{\includegraphics{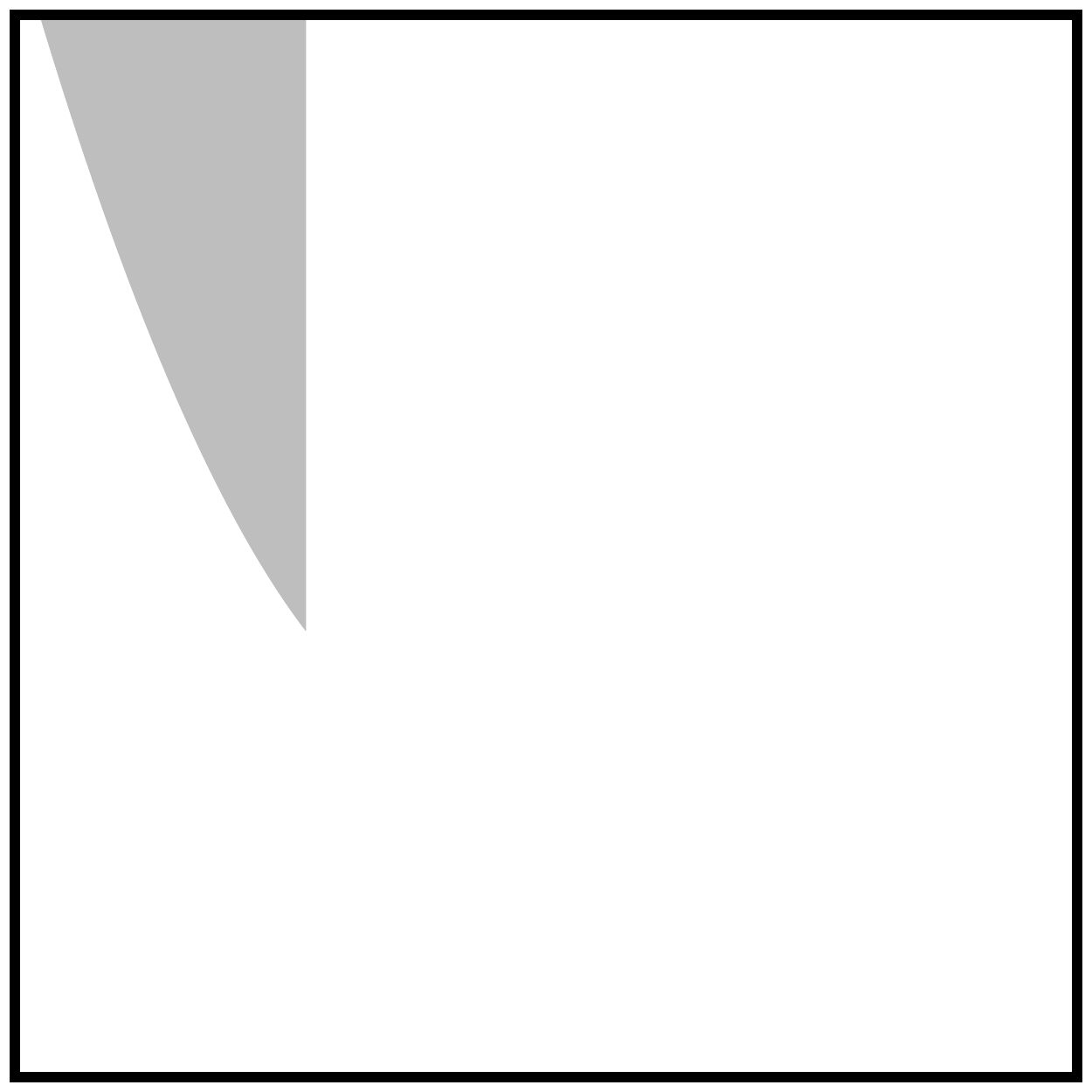}}}
   \item $C_5 = \left(\left[
       \begin{array}{c}
         y>root(f_1,1,y)\myand x > root(f_3,2,x)
         \end{array}
       \right],
lab=2U1U,\alpha=(\frac{3}{2},4),P=\{f_1,f_3\}\right)$ \raisebox{-.4\height}{\scalebox{0.1}{\includegraphics{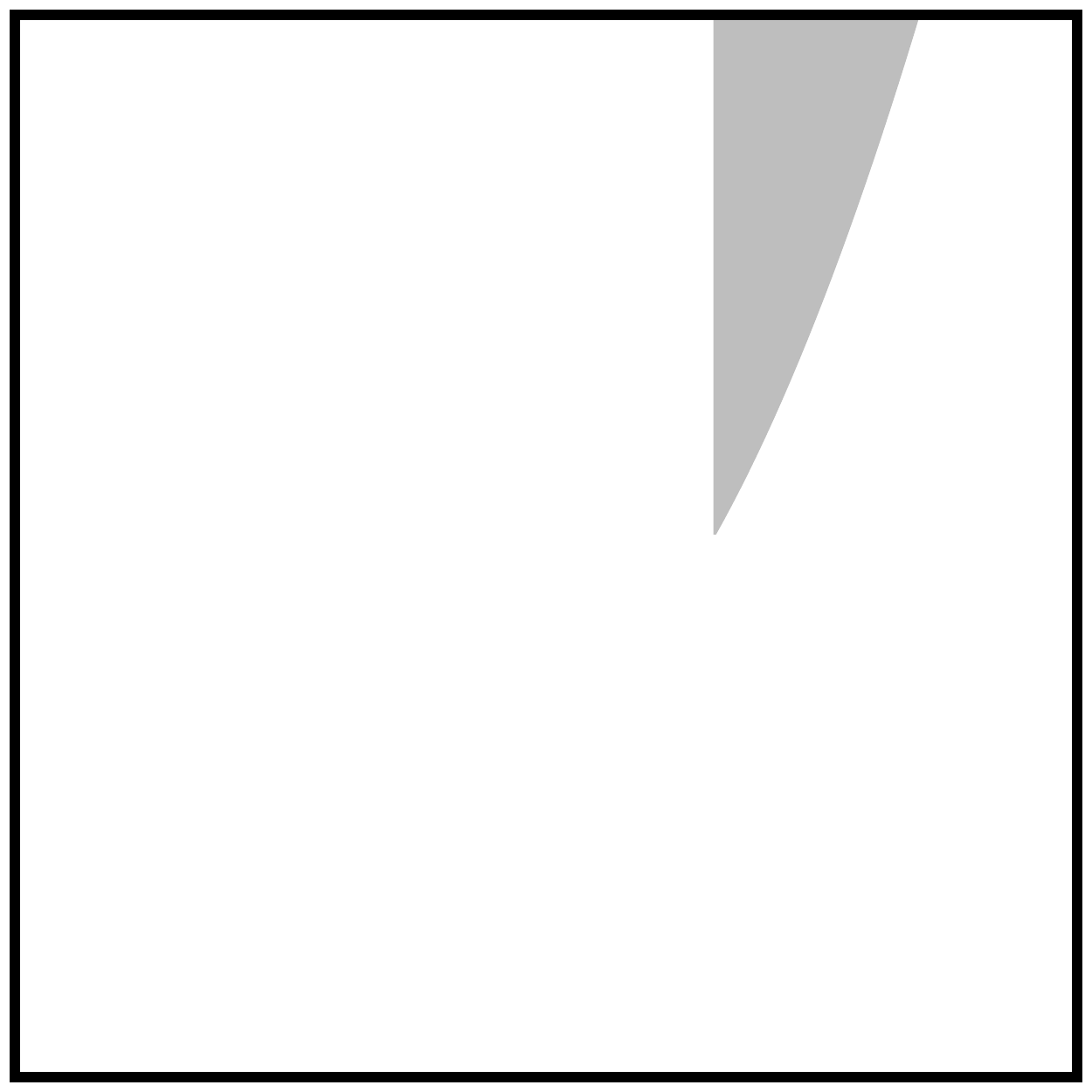}}}
   \item $C_6 = \left(\left[
       \begin{array}{c}
         y > root(f_2,2,y)\myand \\
         x > root(f_3,1,x) \myand x < root(f_3,2,x)
         \end{array}
       \right],
lab=2U2U,\alpha=(0,2),P=\{f_1,f_2^*,f_3,f_4,f_5\}\right)$ \raisebox{-.4\height}{\scalebox{0.1}{\includegraphics{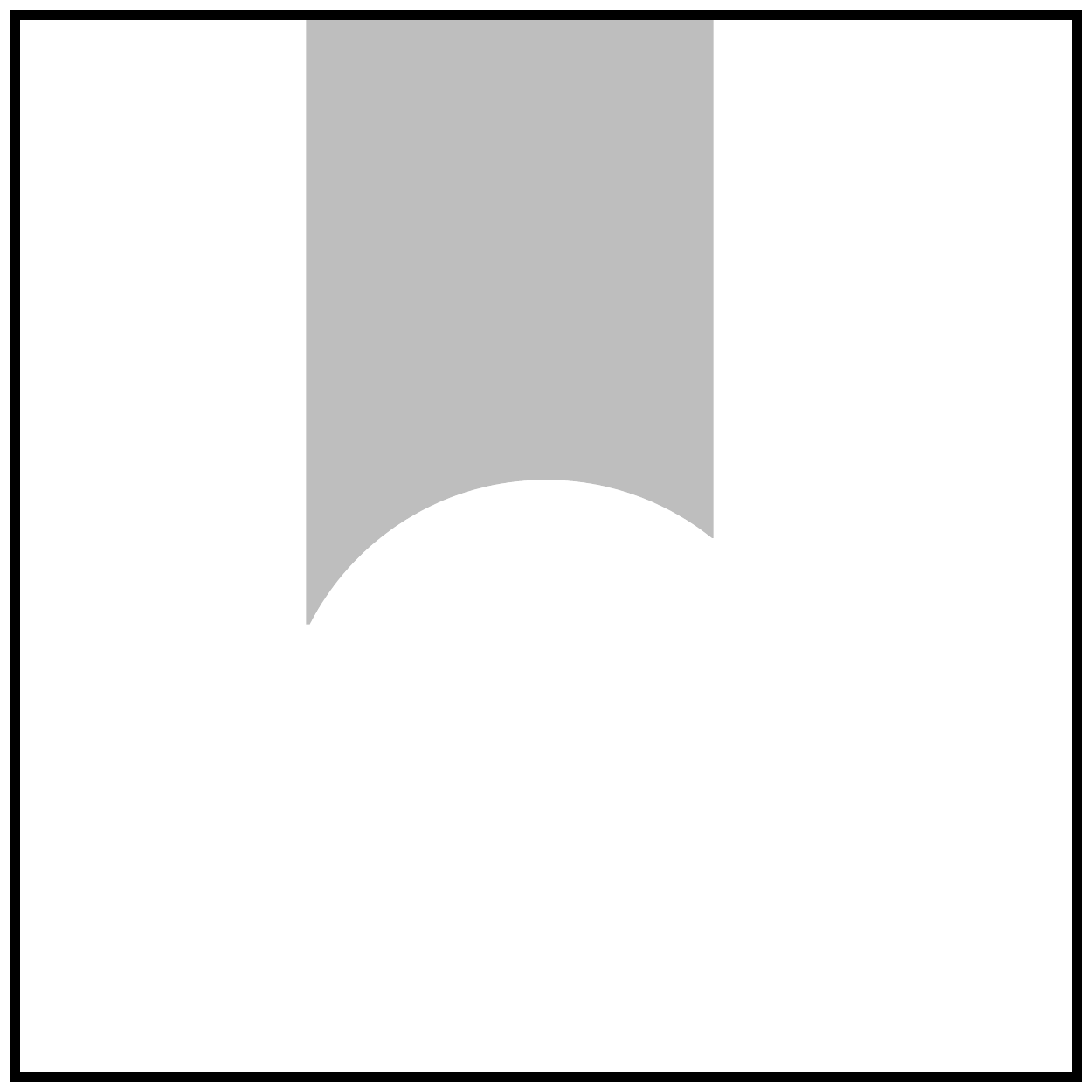}}}
  \end{itemize}
\item $C_3$'s label ends in $X$, so it is not processed further
\item Split($C_4$): $F(\alpha)=\true$, choose $Q = \{f_2\}$, enqueue
  the following
  cells
   \begin{itemize}[leftmargin=0in]
   \item $C_7 = \left(\left[
       \begin{array}{c}
         y>root(f_1,1,y)\myand x < -1
         \end{array}
       \right],
lab=2U1L2X,\alpha=(-\frac{3}{2},2),P=\{f_1,f_2,f_3,f_4,f_5\}\right)$ \raisebox{-.4\height}{\scalebox{0.1}{\includegraphics{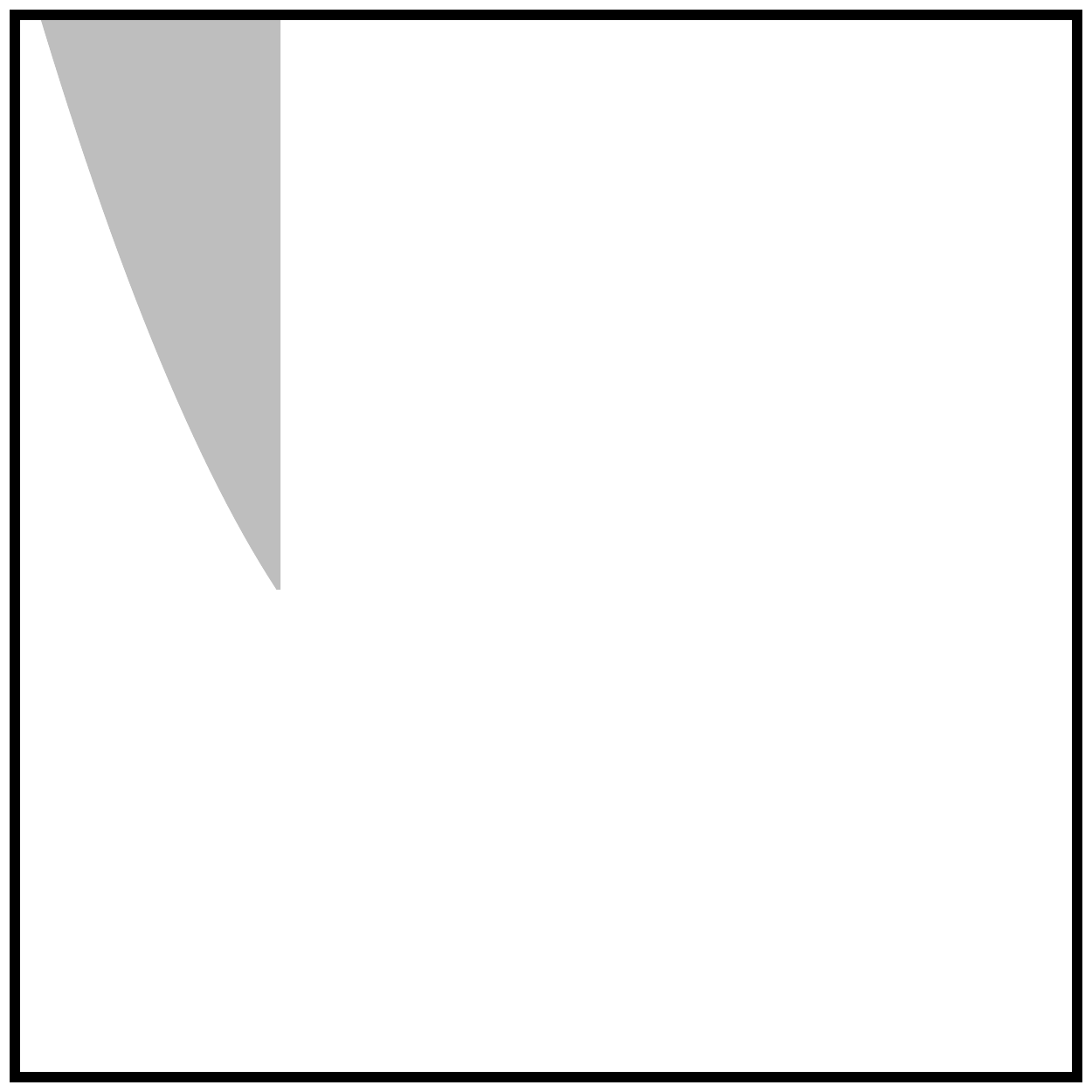}}}
   \item $C_8 = \left(\left[
       \begin{array}{c}
         y>root(f_1,1,y)\myand \\
         x > -1 \myand x < root(f_3,1,x)
         \end{array}
       \right],
lab=2U1L1U,\alpha=(-\frac{15}{16},2),P=\{f_1,f_2,f_3,f_4,f_5\}\right)$ \raisebox{-.4\height}{\scalebox{0.1}{\includegraphics{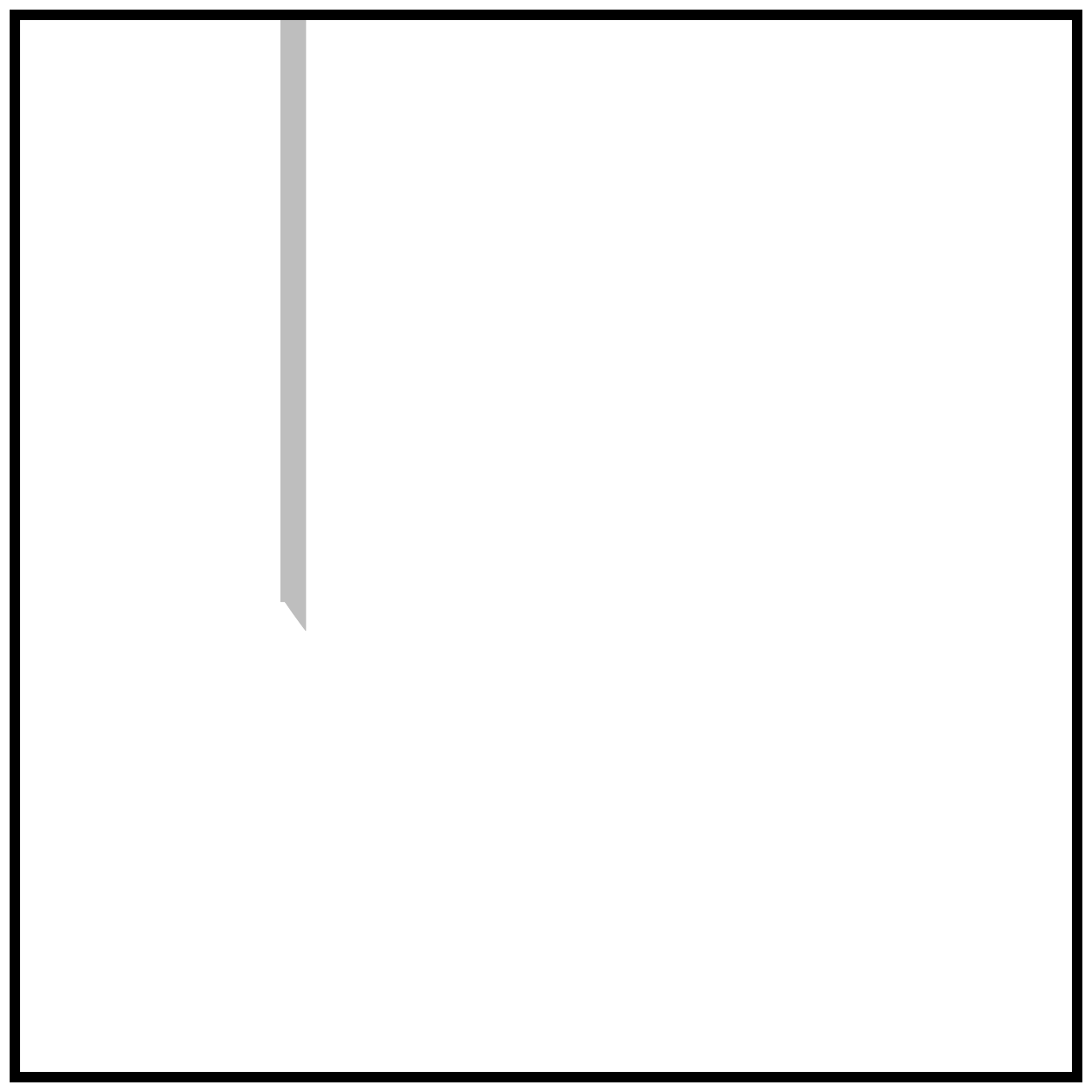}}}
  \end{itemize}
\item Split($C_5$): $F(\alpha)=\true$, choose $Q = \{f_2\}$, enqueue
  the following
  cells
   \begin{itemize}[leftmargin=0in]
   \item $C_9 = \left(\left[
       \begin{array}{c}
         y>root(f_1,1,y)\myand x > 1
         \end{array}
       \right],
lab=2U1U2X,\alpha=(\frac{3}{2},4),P=\{f_1,f_2,f_3,f_4,f_5\}\right)$ \raisebox{-.4\height}{\scalebox{0.1}{\includegraphics{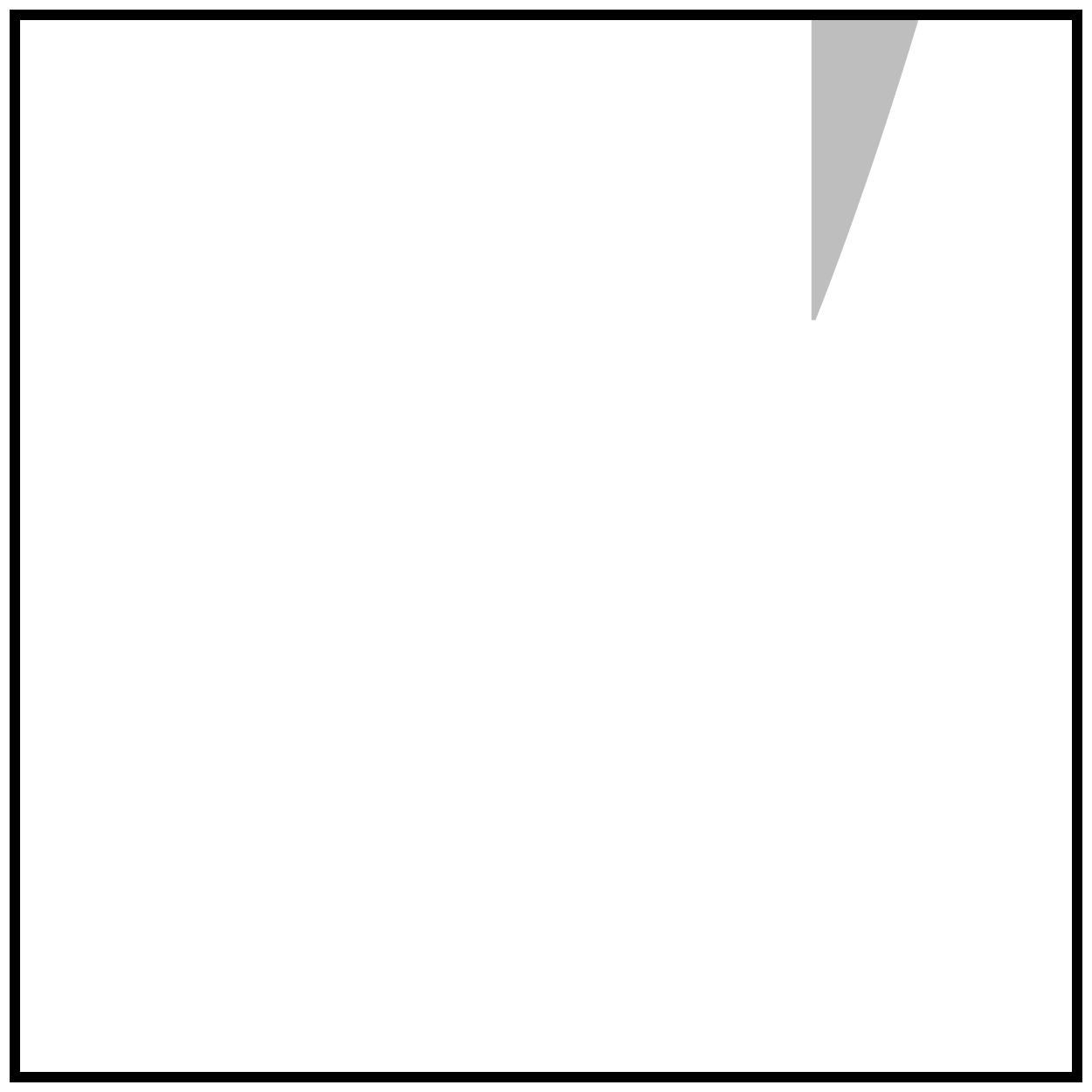}}}
   \item $C_{10} = \left(\left[
       \begin{array}{c}
         y>root(f_1,1,y)\myand \\
         x > root(f_3,2,x)\myand x < 1
         \end{array}
       \right],
lab=2U1U1L,\alpha=(\frac{15}{16},2),P=\{f_1,f_2,f_3,f_4,f_5\}\right)$ \raisebox{-.4\height}{\scalebox{0.1}{\includegraphics{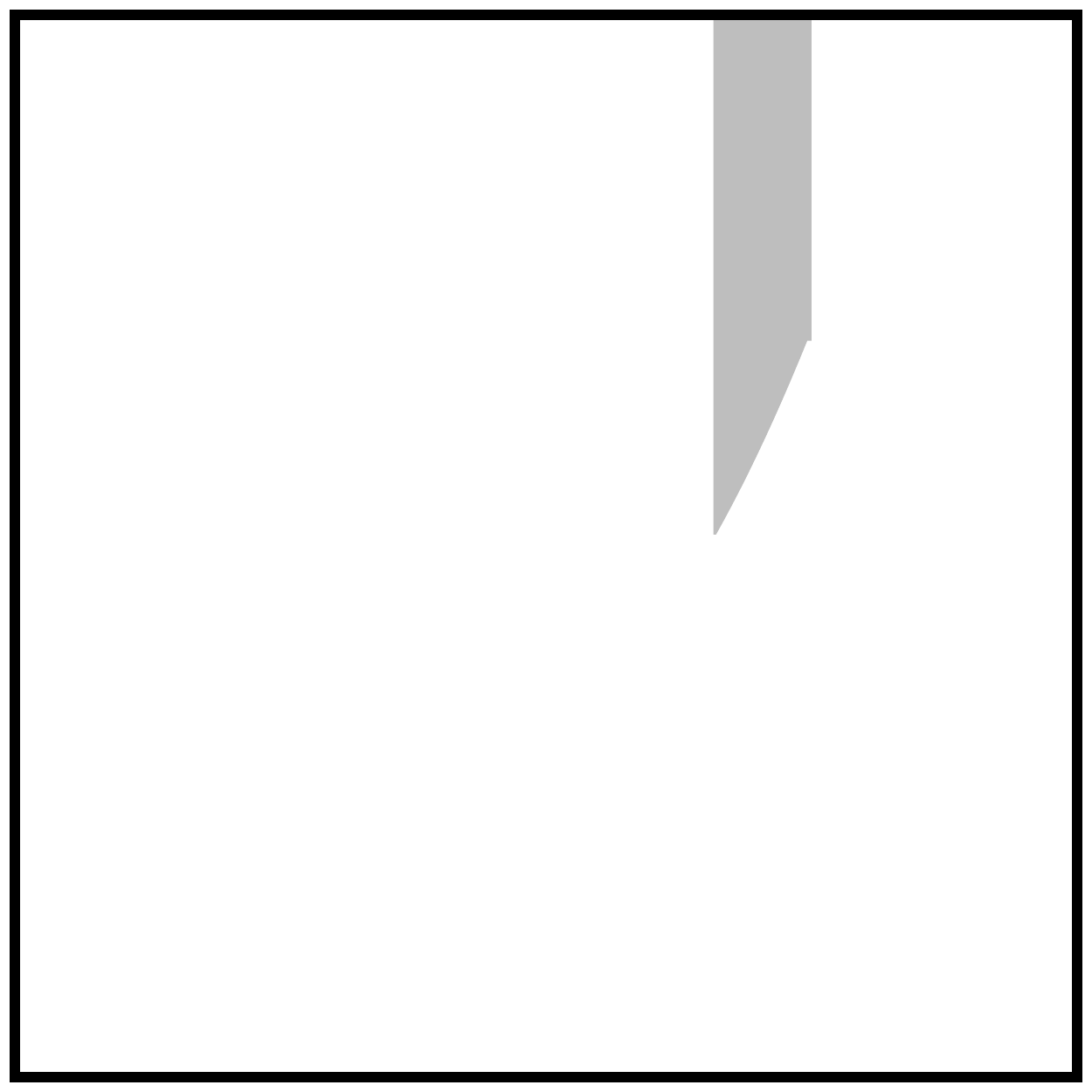}}}
  \end{itemize}
\item all remaining cells in $Q$ either have labels that end in $X$
  or, when the call to Split is made, are not split further.
\end{enumerate}
\begin{figure}
{\scalebox{0.8}{\includegraphics{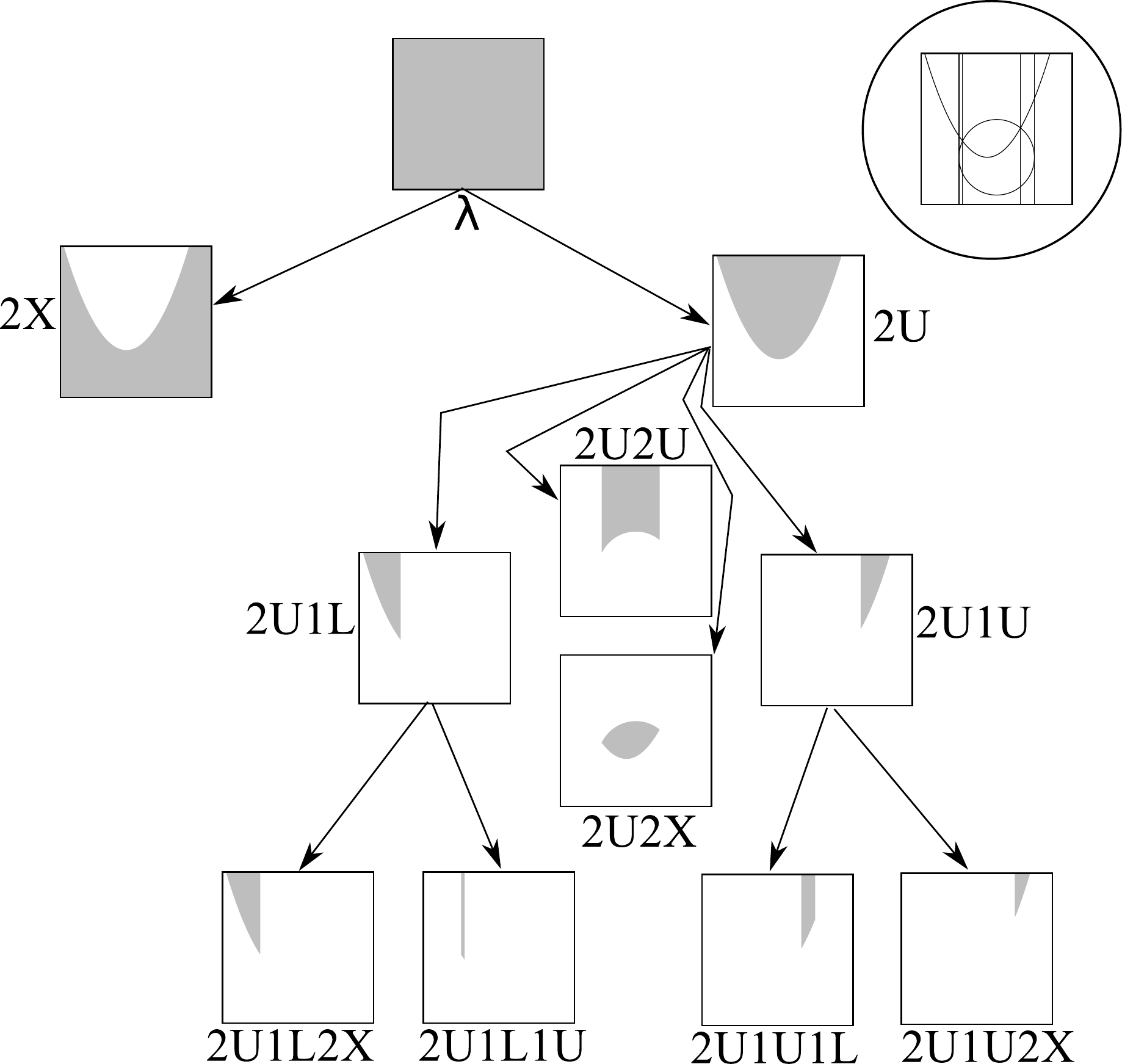}}}
\caption{Depicted here is the Open NuCAD graph structure produced by
  the example run of TI-Open-NuCAD for input formula $F = [ 16 y - 16
  x^2 - 8 x - 1 > 0 \wedge x^2 + y^2 - 1 > 0 ]$.  The leaf nodes are a
  weak decomposition of $\realring^2$ into open cylindrical cells in
  which $F$ is truth-invariant.  Also shown (circled) is the
  truth-invariant CAD for $F$.}
\label{figure:tree}
\end{figure}
\pagebreak
Figure~\ref{figure:tree} shows the NuCAD tree resulting from the above
execution of the TI-Open-NuCAD algorithm.
There are seven leaf nodes, which mean $\realring^2$ has been
decomposed into seven cells.  The standard truth-invariant CAD for
input formula $F$ (shown circled in Figure~\ref{figure:tree})
contains 16 open cells in $\realring^2$.  The Open NuCAD fails to be
an Open CAD because the projections onto $\realring^1$ of 
the cell $2X$ and any other leaf cell are neither disjoint nor
identical. 

The primary purpose of this example is to illustrate the basic
functioning of TI-Open-NuCAD, and to illustrate the Open NuCAD data
structure.  Hopefully it has been successful in this.  There are
two important limitations to this example, though.  First of all, 
Step~3, which deals with ``fail'' results returned by the 
OC-Merge-Set operation, is not illustrated.  Secondly, and more
importantly, because this example only involves two variables there 
is no opportunity to illustrate the reduction in the number and size
of projection factor sets that we expect to accompany the model-based
approach to CAD construction.

\section{The correctness of TI-Open-NuCAD}
In this section we sketch a proof of the correctness of 
TI-Open-NuCAD.  In fact, TI-Open-NuCAD clearly 
meets its specification provided that Split meets its specification,
and that termination can be proved.  First we prove a lemma that is
key to showing the termination of TI-Open-NuCAD.

For Open OneCell $D$ we
denote the set of polynomials whose sections define the boundaries
of $D$ by $\mbox{bpolys}(D)$ (note that they will be irreducible).  For
Tarski formula $F$ we denote the 
set of irreducible factors of polynomials appearing on the
left-hand-side of the atomic
formulas of $F$ when they are normalized to be of the form 
$f\ \sigma \ 0$ by $factors(F)$.

\begin{lemma}
\label{lemma:factors}
Suppose the call $\mbox{Split}(D,F)$ produces a non-empty queue $Q'$.
Let $H$ be the closure under the Open McCallum projection of
$\mbox{bpolys}(D) \cup factors(F)$.
For each cell $C \in Q'$, $\mbox{bpolys}(C) \subseteq H$.
\end{lemma}
\proof
First we note that if Step~2 produces $D' = (\fail,f)$ then although 
the sample point $\alpha$ and some of the algebraic numbers in the
data-structure may change, the defining formula for $D'$ and,
therefore, the elements of $\mbox{bpolys}(D)$ remain the same.
Next we note that if Step~2 produces a cell $D'$ (i.e. does not
produce $\fail$) then the specification of the O-P-Merge algorithm
from \cite{Brown:2013}, and by extension the OC-Merge-Set algorithm
called in Step~2, guarantees that $\mbox{bpolys}(D')$ is a subset
of the closure under the Open McCallum projection of 
$\mbox{bpolys}(D) \cup Q$.  Since $Q \subseteq factors(F)$, we have
$\mbox{bpolys}(D') \subseteq H$.
For any cell $C$ enqueued on the output queue, at each level $i$,
the boundaries of $C$ are sections of polynomials from the set 
$\{D[i].l,D'[i].l,D'[i].u,D[i].u\}$, which is a subset of $H$.
\qed

\begin{lemma}
\label{lemma:split}
The Algorithm $\mbox{Split}(D,F)$ terminates and meets its
specification.
\end{lemma}
\proof
As long as Step~3 only produces new values for point $\alpha$ that
are in the cell defined by $D$ and Step~2 eventually produces a non-$\fail$
result, $\mbox{Split}(D,F)$ clearly meets it specification.
Moreover, if the body of Step~3 is executed and $\alpha$ is in the
cell defined by $D$ (which is certainly true initially), then the new
value of $\alpha$ is also in the cell defined by $D$.  This is clear
because $\gamma_i$ is chosen from the interval
$(\mbox{max}(\zeta,D[i].L),\alpha_i) \subset (D[i].L,D[i].U)$,
and for $j \in \{i+1,\ldots,n\}$, $\gamma_j$ is chosen specifically
to satisfy the defining formula
$$root(D[j].l,D[j].L.j,x_j) < x_j < root(D[j].u,D[j].U.j,x_j).$$
What remains to be proven is termination, which boils down to showing
that the call to OC-Merge-Set in Step~2 eventually returns a
non-$\fail$ result.  If we were assured that OC-Merge-Set would
produce the same projection factors for the perturbed $\alpha$ as for
the original, this would be clear.  Unfortunately, we cannot be sure
of that.  Thus, we require a more subtle argument.  First, we note
that each perturbation leaves the $x_k$th coordinate unchanged for all
$k < i$, reduces the $i$th coordinates $\alpha_i$ so that it changes
from a root of $g(\alpha_1,\ldots,\alpha_{i-1},x_i)$ to something
slightly smaller (Step 3c), and potentially changes the remaining
coordinates.

Suppose Split does not terminate.  Then there is an infinite sequence 
of $\alpha$ values and associated $f$'s satisfying $f(\alpha) = 0$.
Note that all the polynomials $f$ as well as all the elements of the
set $L$ constructed from $F$ come from the closure under the
McCallum projection of $\mbox{bpolys}(D) \cup \mbox{factors}(F)$,
which we'll denote $P_{MC}$. Let
$\alpha^{(0)}, \alpha^{(1)},\ldots$ be the infinite sequence of values
for $\alpha$ as the process progresses, let 
$f^{(0)}, f^{(1)},\ldots$ be the infinite sequence of associated $f$'s
and $L^{(0)}, L^{(1)},\ldots$ and $i^{(0)}, i^{(1)},\ldots$
be the associated values for $L$ and $i$ arrived at by Step 3b.  
We note that for any $k$, the elements $L^{(k)}$ all divide
$A(i^{(k)},\alpha^{(k)})$, where
$$
A(m,\rho) = \prod_{g \in P_{MC} |level(g) = m \wedge
  g(\rho_1,\ldots,\rho_{m-1},x_m) \neq 0}g.$$
We also note that the polynomial set $\{A(m,\rho)|m\in\{1,\ldots,n\} \wedge \rho 
\in \realring^n\}$ is finite.
So, for each $k$ we have that that $i^{(k)}$th coordinate of
$\alpha^{(k)}$ is a zero of some $g \in L^{(k)}$ that is not
nullified at $(\alpha_1^{(k)},\ldots,\alpha_{i^{(k)}-1}^{(k)})$
and thus is a zero of $A(i^{(k)},\alpha^{(k)})$.

We will show that for each level $r$, there is a value $k$ after which 
the $r$th coordinate of $\alpha^{(k)}$ never changes.  We proceed by
induction on $r$.

Consider the case $r=1$.  Consider the subsequence $k_1,k_2,\ldots$ 
of all indices $k$ for
which $i^{(k)} = 1$.  For each $k_j$ in this subsequence,
$\alpha_1^{(k_j)}$ is a zero of $A(1,\alpha^{(k_j)})$.
Moreover, the new 
value of $\alpha_1$ is smaller than the previous value and, since the
value of the 1st component of $\alpha$ is otherwise never changed, 
$\alpha_1^{(k)}$
is strictly decreasing over the subsequence $k_1, k_2, \ldots$.  Since 
$A(1,\beta)$ is the same for any $\beta\in\realring^n$, and it has
finitely many roots, there are only finitely many elements of the
subsequence. In particular, there is a largest index $k^*$ in the
subsequence ($k^*$ can be taken as zero if the subsequence is empty),
and $\alpha_1$ is constant over all indices greater than $k^*$. 

Suppose $r > 1$.  Assume, by induction, that the result holds for all
smaller values of $r$.  Then there is an index $k'$ such that for all
$k > k'$ the first $r-1$ components of $\alpha^{(k)}$ are constant.
So, for all $k > k'$, the $r$th component of $\alpha^{(k)}$ is
non-increasing.
Consider the subsequence $k_1,k_2,\ldots$ 
of all indices $k > k'$ for
which $i^{(k)} = r$.  
Note that because the $r$th component of $\alpha$ is reduced at each step for
which $i^{(k)} = r$,
the sequence of values $\alpha_r^{(k_1)},\alpha_r^{(k_2)},\ldots$
is strictly decreasing.
For each $k_j$ in the subsequence,
$\alpha_1^{(k_j)}$ is a zero of $A(r,\alpha^{(k_j)})$.
Since there are only finitely many polynomials $A(r,\beta)$, 
where $\beta \in
(\alpha_1^{(k'+1)},\ldots,\alpha_{r-1}^{(k'+1)})\times\realring^{n-r+1}$,
each having only finitely many roots, there are only finitely many
elements in the subsequence.  In particular, there is a largest index
$k^*$ in the 
subsequence ($k^*$ can be taken as $k'$ if the subsequence is empty),
and $\alpha_r$ is constant over all indices larger than $k^*$. 

Thus, we have proven that there is an index $k'$  such that
for all $k > k'$, all coordinates of $\alpha^{(k)}$ are constant.
This is a contradiction, since executing Step 3 changes $\alpha$,
which means that our assumption that there is an input for which Split
does not terminate is invalid. 
This completes our proof of the termination and correctness of Split. 
\qed

\begin{theorem}
Algorithm TI-Open-NuCAD terminates, and meets its specification.
\end{theorem}
\proof
Lemma~\ref{lemma:split} shows that Split terminates and is correct.
Lemma~\ref{lemma:factors} shows that the boundary polynomials for
the cells returned by Split are elements of the closure
under the Open McCallum projection of $\mbox{factors}(F)$.
Thus for any cell $D'$ returned by
Split, and any cell $C$ from the
CAD produced by the Open McCallum projection for $F$, either
$C \cap D' = \emptyset$ or $C \subseteq D'$.  This means that
for each each cell $D$ enqueued on $Q$, we can imagine associating
with $D$ the
set of cells from the
CAD produced by the Open McCallum projection for $F$ that are
contained in $D$ --- we call this set $M_D$.  
Note that $M_D$ is never empty.
Recall that when a cell with label ending in $X$ is dequeued from $Q$,
no call to 
Split is made.  
Define $X_Q$ to be the set of cells in $Q$ with label ending in $X$.
Consider the quantity
\begin{equation}
\label{equation:cq}
c_Q = |X_Q| + \sum_{E\in Q - X_Q}2|M_E|^2.
\end{equation}
We will show 
 that at each iteration of the loop in Step 4 of
TI-Open-NuCAD the quantity $c_Q$ is reduced.
Every iteration, a cell $D$
is dequeued from $Q$ and 
one of the following occurs:
\begin{enumerate}
\item no new cells are enqueued --- in which case one of the terms
on the right-hand side of (\ref{equation:cq}) gets smaller and the
other term is unchanged,
\item a single cell whose label ends in $X$ is enqueued --- in which
  case $|X_Q|$ increases by one, but 
  $\sum_{E\in Q - X_Q}2|M_E|^2$ is reduced by $2|M_D|^2 > 1$ 
, 
\item more than one cell is enqueued --- in which case
  the $|X_Q|$ term is increased by one, but in the sum
  the term $2|M_D|^2$ is replaced by 
  $2|M_{D_1}|^2 + 2|M_{D_2}|^2 + \cdots + 2|M_{D_t}|^2$ where
  $|M_D| = |M_{D_1}| + |M_{D_2}| + \cdots + |M_{D_t}|$, $t \geq 2$. 
  So the net change is 
$$
1 + 2|M_{D_1}|^2 + 2|M_{D_2}|^2 + \cdots + 2|M_{D_t}|^2 - 2|M_D|^2 < 0.
$$
\end{enumerate}
Thus, termination is proven and, as noted
previously, correctness is then easily verified.
\qed

\section{Advantages of the model-based approach}
Further work is required to either produce an implementation of these
algorithms and provide a systematic empirical comparison between them
and the usual Open CAD construction algorithm, or to provide an
analytical comparison.  Moreover, in as much as an Open NuCAD is less
structured than an Open CAD, it cannot necessarily be used for the
same purposes.  So yet more work is required to understand the
applications and limitations of this new variant of CAD.  Given these
points, it is worth listing some of the reasons why the model-based
approach and Open NuCADs are important and worth developing.
\begin{enumerate}
\item The model-based approach produces smaller projection-factor sets
  and larger sign-invariant cells.  This point is demonstrated in
  \cite{Brown:2013}, and further experiments showing this were
  presented in the ISSAC 2013 talk accompanying that paper.  This is
  perhaps the most important reason to pursue this new approach,
  because the reduction in the number of projection factors and the
  increase in cell size is substantial.  For a single cell,
  experiments point to exponentially smaller projection factor sets
  and exponentially larger cells.
\item NuCADs allow for truth-invariant decompositions for an input
  formula using fewer cells than CADs.  The example in this paper 
  demonstrates this point, although certainly more analysis, either
  empirical or analytical, is required to understand how substantial
  the difference between NuCADs and CADs really is.
\item Model-based construction of NuCADs is incremental.  After one
  loop iteration, which requires a small amount of time and space relative to
  even just the projection step for CAD construction, the new approach
  produces a cell in which the input formula is truth-invariant.  This
  is in marked contrast with traditional CAD construction, for which 
  the entire projection must be computed before even the first cell is
  constructed. 
\item Model-based construction of NuCADs is naturally parallelizable.
  Splitting of one cell in the queue $Q$ is completely independent of 
  splitting other cells, so all the splitting can be done in
  parallel.  In fact, nodes could keep their own queues of cells to 
  split, and would only need to communicate when one node ran out of
  cells to split and had to steal some from another's queue.  The one
  kind of information that one would probably want nodes to share 
  would be the results of particularly expensive resultant and
  discriminant computations and the accompanying factorizations.
\end{enumerate}

Perhaps the most interesting of all, however, is that none of the
proofs of the doubly-exponential worst-case running time of CAD apply
to NuCADs.
\cite{BrownDavenport:07,DavenportHeintz:88,Weispfenning:88}
all deduce the doubly-exponential worst-case performance of CAD
from its connection to quantifier elimination --- in particular, 
quantifier elimination for formulas with many quantifier alternations.
NuCADs, however, do not directly allow for quantifier elimination, at
least not for formulas with quantifier alternations, so they are not
subject to that argument.  This leaves open the intriguing possibility
that the model-based approach and NuCADs may provide a CAD-style 
algorithm for satisfiability, existential quantifier elimination, or
even full quantifier elimination with an asymptotic complexity 
competitive with modern QE algorithms, but with the kind of practical
utility that has made CAD attractive for smaller problems.

\bibliographystyle{plain}
\bibliography{BibEntries}

\end{document}